\documentclass[aps,preprint,amsmath,amssymb,showpacs]{revtex4}

\usepackage{graphicx}
\begin{document}

\title{Anomalous quartic $WW\gamma \gamma$ and $WWZ\gamma$ couplings through
$W^{+}W^{-}Z$ production in $\gamma\gamma$ colliders}

\author{\.{I}nan\c{c} \c{S}ahin}
\email[]{isahin@wisc.edu} \email[]{isahin@science.ankara.edu.tr}
\affiliation{Department of Physics, University of Wisconsin,
Madison, WI 53706, USA}
 \affiliation{Department of
Physics, Faculty of Sciences, Ankara University, 06100 Tandogan,
Ankara, Turkey}

\begin{abstract}
We find 95\% confidence level limits on the anomalous coupling
parameters $a_0$, $a_c$ and $a_n$ with an integrated luminosity of
$500fb^{-1}$ and $\sqrt s =0.5,\,1$ and 1.5 TeV energies. We take
into account incoming beam polarizations and also the final state
polarizations of the gauge bosons in the cross-section calculations
to improve the bounds. We show that polarization leads to a
significant amount of improvement in the sensitivity limits.
\end{abstract}

\pacs{12.15.Ji, 12.60.Cn, 13.88.+e}

\maketitle

\section{Introduction}

Gauge boson self interactions are strictly constrained by the
$SU_{L}(2)\times U_{Y}(1)$ gauge invariance.  Therefore the direct
study of gauge boson self interactions provide a crucial test of the
gauge structure of the standard model (SM). Any deviation of the
couplings from the expected values would indicate the existence of
new physics beyond the SM. In the recent experiments at CERN
$e^+e^-$ collider LEP and Fermilab Tevatron gauge boson self
interactions have been studied experimentally through $e^+e^-,p\bar
p \to W^+ W^-, WZ, ZZ, Z\gamma, W^+ W^- \gamma, Z\gamma \gamma$
gauge boson production processes
\cite{LEP,LEP2,LEP3,Tevatron1,Tevatron2}. It will be possible to
produce a final state with three or more massive gauge bosons in the
next generation of $e^+e^-$ colliders. After these future $e^+e^-$
colliders are constructed its operating modes of $e\gamma$ and
$\gamma\gamma$ are expected to be designed \cite{akerlof}.

Future $e^+e^-$ collider and its $e\gamma$ and $\gamma\gamma$ modes
will have a great potential to probe anomalous quartic gauge boson
vertices. We concentrate on genuine quartic gauge boson couplings
which do not induce new trilinear vertices. Genuine quartic
couplings are contact interactions, manifestations of the exchange
of heavy particles. On the other hand non-genuine quartic gauge
boson couplings emerge from an operator that induces both trilinear
and quartic gauge boson couplings. Non-genuine couplings can be
investigated much more efficiently through their trilinear
counterpart. In this paper we assume that quartic couplings are
modified by genuine anomalous interactions while the trilinear
couplings are all given by their SM values.

In writing effective operators associated with genuinely quartic
couplings we employ the formalism of \cite{eboli1}. Imposing
custodial $SU(2)_{Weak}$ symmetry and local $U(1)_{em}$ symmetry and
if we restrict ourselves to C and P conserving interactions,
dimension 6 effective lagrangian for $W^+W^-\gamma\gamma$,
$ZZ\gamma\gamma$ and $W^{+}W^{-}Z\gamma$ couplings are given by,

\begin{eqnarray}
{\cal L}&&={\cal L}_0+{\cal L}_c+{\cal L}_n
\end{eqnarray}

\begin{eqnarray}
{\cal
L}_{0}&&=\frac{-\pi\alpha}{4\Lambda^{2}}a_{0}F_{\mu\nu}F^{\mu\nu}
W_{\alpha}^{(i)}W^{(i) \alpha}
\end{eqnarray}

\begin{eqnarray}
{\cal
L}_{c}&&=\frac{-\pi\alpha}{4\Lambda^{2}}a_{c}F_{\mu\alpha}F^{\mu\beta}
W^{(i) \alpha}W_{\beta}^{(i)}
\end{eqnarray}

\begin{eqnarray}
{\cal L}_{n}&&=\frac{i\pi\alpha}{4\Lambda^{2}}a_{n}\epsilon_{ijk}
W_{\mu\alpha}^{(i)}W_{\nu}^{(j)}W^{(k)\alpha}F^{\mu\nu}
\end{eqnarray}
where $W^{(i)}$ is the $SU(2)_{Weak}$ triplet, and $F_{\mu\nu}$ and
$W_{\mu\alpha}^{(i)}$ are the electromagnetic and $SU(2)_{Weak}$
field strengths respectively. $a_0$, $a_c$ and $a_{n}$ are the
dimensionless anomalous coupling constants. Effective lagrangians
(2) and (3) give rise to anomalous $W^+W^-\gamma\gamma$ and also
$ZZ\gamma\gamma$ couplings. Effective lagrangian (4) give rise to
$W^{+}W^{-}Z\gamma$ coupling. For sensitivity calculations to the
anomalous couplings we set the new physics energy scale $\Lambda$ to
$M_{W}$.

The vertex functions for
$W^{+}(k_{1}^{\mu})W^{-}(k_{2}^{\nu})\gamma(p_{1}^{\alpha})
\gamma(p_{2}^{\beta})$  and
$W^{+}(p_{+}^{\mu})W^{-}(p_{-}^{\nu})Z(p_{1}^{\alpha})
\gamma(p_{2}^{\beta})$  generated from the effective lagrangians
(2), (3) and (4) are given respectively by

\begin{eqnarray}
i\frac{2\pi\alpha}{\Lambda^{2}}a_{0}g_{\mu\nu}\left[g_{\alpha\beta}
(p_{1}.p_{2})-p_{2 \alpha}p_{1 \beta}\right]
\end{eqnarray}

\begin{eqnarray}
i\frac{\pi\alpha}{2\Lambda^{2}}a_{c}\left[(p_{1}.p_{2})(g_{\mu\alpha}
g_{\nu\beta}+g_{\mu\beta}g_{\alpha\nu})+g_{\alpha\beta}(p_{1\mu}p_{2\nu}
+p_{2\mu}p_{1\nu})\right. \nonumber \\
\left. -p_{1\beta}(g_{\alpha\mu}p_{2\nu}+g_{\alpha\nu}p_{2\mu})
-p_{2\alpha}(g_{\beta\mu}p_{1\nu}+g_{\beta\nu}p_{1\mu})\right]
\end{eqnarray}

\begin{eqnarray}
i\frac{\pi\alpha}{4\cos\theta_{W}\Lambda^{2}}a_{n}\left[g_{\mu\alpha}\left[g_{\nu\beta}
p_{2}.(p_{1}-p_{+})-p_{2\nu}(p_{1}-p_{+})_{\beta}\right] \right.
\nonumber \\ \left.
-g_{\nu\alpha}\left[g_{\mu\beta}p_{2}.(p_{1}-p_{-})-p_{2\mu}(p_{1}-p_{-})_{\beta}\right]
\right. \nonumber \\ \left.
+g_{\mu\nu}\left[g_{\alpha\beta}p_{2}.(p_{+}-p_{-})-p_{2\alpha}(p_{+}-p_{-})_{\beta}\right]
 \right. \nonumber \\ \left. -p_{1\mu}(g_{\nu\beta} p_{2\alpha}-g_{\alpha\beta} p_{2\nu})
 +p_{1\nu}(g_{\mu\beta} p_{2\alpha}-g_{\alpha\beta} p_{2\mu})\right. \nonumber \\ \left.
-p_{-\alpha}(g_{\mu\beta} p_{2\nu}-g_{\nu\beta}p_{2\mu})
+p_{+\alpha}(g_{\nu\beta} p_{2\mu}-g_{\mu\beta}
p_{2\nu})\right.\nonumber \\ \left. -p_{+\nu}(g_{\alpha\beta}
p_{2\mu}-g_{\mu\beta}p_{2\alpha}) +p_{-\mu}(g_{\alpha\beta}
p_{2\nu}-g_{\nu\beta} p_{2\alpha})
 \right]
\end{eqnarray}
For a convention, we assume that all the momenta are incoming to the
vertex. The anomalous $ZZ\gamma\gamma$ couplings are obtained by
multiplying (5), (6) by $\frac{1}{\cos^{2}\theta_{W}}$ and making $W
\to Z$.

CERN $e^{+}e^{-}$ collider LEP provide present collider limits on
anomalous quartic $W^+W^-\gamma\gamma$ and $W^{+}W^{-}Z\gamma$
couplings. Recent results from OPAL collaboration for
$W^{+}W^{-}\gamma\gamma$ couplings are given by -0.020 $GeV^{-2} <
\frac{a_{0}}{\Lambda^{2}} <$ 0.020 $GeV^{-2}$, -0.052 $GeV^{-2} <
\frac{a_{c}}{\Lambda^{2}} <$ 0.037 $GeV^{-2}$ at 95\% C.L. assuming
that the $W^{+}W^{-}\gamma\gamma$ couplings are independent of the
$ZZ\gamma\gamma$ couplings \cite{LEP3}. If it is assumed that
$W^{+}W^{-}\gamma\gamma$ couplings are dependent on the
$ZZ\gamma\gamma$ couplings then the 95\% C.L. sensitivity limits are
improved to -0.002 $GeV^{-2} < \frac{a_{0}}{\Lambda^{2}} <$ 0.019
$GeV^{-2}$, -0.022 $GeV^{-2} < \frac{a_{c}}{\Lambda^{2}} <$ 0.029
$GeV^{-2}$. Recent results from L3, OPAL and DELPHI collaborations
for $W^{+}W^{-}Z\gamma$ coupling are given by -0.14 $GeV^{-2} <
\frac{a_{n}}{\Lambda^{2}} <$ 0.13 $GeV^{-2}$, -0.16 $GeV^{-2} <
\frac{a_{n}}{\Lambda^{2}} <$ 0.15 $GeV^{-2}$ and -0.18 $GeV^{-2} <
\frac{a_{n}}{\Lambda^{2}} <$ 0.14 $GeV^{-2}$ at 95\% C.L.
respectively \cite{LEP,LEP2,LEP3}.

In the literature there has been a great amount of work on anomalous
quartic $W^+W^-\gamma\gamma$ and $W^{+}W^{-}Z\gamma$ couplings which
focus on future linear $e^+e^-$ collider and its $e\gamma$ and
$\gamma\gamma$ modes. Anomalous quartic $W^+W^-\gamma\gamma$ and
$W^{+}W^{-}Z\gamma$ couplings have been studied through the
reactions $e^{+}e^{-} \to VVV$ \cite{barger}, $e^{+}e^{-} \to FFVV$
\cite{boos}, $e\gamma \to VVF$ \cite{eboli1,isahin1,isahin2},
$\gamma\gamma \to VV$ \cite{boudjema}, $\gamma\gamma \to VVV$
\cite{eboli2} and $\gamma\gamma \to VVVV$ \cite{eboli3} where
$V=Z,W$ or $\gamma$ and $F=e$ or $\nu$. These couplings have also
been studied at hadron colliders through the reactions
$pp(\bar{p})\to \gamma\gamma Z$, $\gamma\gamma W$ and $qq \to
qq\gamma\gamma$, $qq\gamma Z$ \cite{eboli4}.

In this work we have investigated anomalous quartic $W^+W^-\gamma
\gamma$ and $W^+W^-Z\gamma$ couplings via the process $\gamma \gamma
\to W^{+}W^{-}Z$. This process involve only interactions between the
gauge bosons, making more evident any deviation from the SM
predictions. It does not contains any tree-level Higgs contribution.
Therefore it excludes all the uncertainties coming from the scalar
sector, such as the Higgs boson mass. The same process was analyzed
in \cite{eboli2} with unpolarized initial and final states. We take
into account incoming beam polarizations and also the final state
polarizations of the gauge bosons in the cross-section calculations
to improve the bounds. We have showed that polarization leads to a
significant amount of improvement in the sensitivity limits.

\section{Cross sections for polarized beams}

The process $\gamma \gamma \to W^{+}W^{-}Z$ is described by twelve
tree-level SM diagrams and two new diagrams that consist of an
anomalous vertex $ZZ\gamma\gamma$ and a five-vertex $\gamma\gamma
W^+ W^- Z$ which is necessary to preserve the gauge invariance of
the amplitude \cite{eboli2}. Feynman rules for this five-vertex is
given in Ref.\cite{eboli2}. Since $W^+W^-\gamma \gamma$ and
$W^+W^-Z\gamma$ couplings are non-zero in the SM, they contribute to
the SM diagrams and modify SM amplitudes. The analytical expression
for the cross section is quite lengthy and we have evaluated
numerically. The phase space integrations have been performed by
GRACE \cite{grace} which uses a Monte Carlo routine. As a check of
our results we have confirmed the results of \cite{eboli2} for
unpolarized beams.

The most promising mechanism to generate energetic gamma beams in an
$e^{+}e^{-}$ linear collider is Compton backscattering. The spectrum
of Compton backscattered photons is given by \cite{Ginzburg}

\begin{eqnarray}
f_{\gamma/e}(y)={{1}\over{g(\zeta)}}[1-y+{{1}\over{1-y}}
-{{4y}\over{\zeta(1-y)}}+{{4y^{2}}\over {\zeta^{2}(1-y)^{2}}}+
\lambda_{0}\lambda_{e} r\zeta (1-2r)(2-y)]
\end{eqnarray}

where

\begin{eqnarray}
g(\zeta)=&&g_{1}(\zeta)+
\lambda_{0}\lambda_{e}g_{2}(\zeta) \nonumber\\
g_{1}(\zeta)=&&(1-{{4}\over{\zeta}}
-{{8}\over{\zeta^{2}}})\ln{(\zeta+1)}
+{{1}\over{2}}+{{8}\over{\zeta}}-{{1}\over{2(\zeta+1)^{2}}} \\
g_{2}(\zeta)=&&(1+{{2}\over{\zeta}})\ln{(\zeta+1)}
-{{5}\over{2}}+{{1}\over{\zeta+1}}-{{1}\over{2(\zeta+1)^{2}}}
\end{eqnarray}
with $r=y/[\zeta(1-y)]$ and $\zeta=4E_{e}E_{0}/M_{e}^{2}$. $E_{0}$
and $\lambda_{0}$ are the energy and helicity of initial laser
photon and $E_{e}$ and $\lambda_{e}$ are the energy and the helicity
of initial electron beam before Compton backscattering. $y$ is the
fraction which represents the ratio between the scattered photon and
initial electron energy for the backscattered photons moving along
the initial electron direction. Maximum value of $y$ reaches 0.83
when $\zeta=4.8$ in which the backscattered photon energy is
maximized without spoiling the luminosity.

Backscattered photons are not in fixed helicity states their
helicities are described by a distribution \cite{Ginzburg}:

\begin{eqnarray}
\xi(E_{\gamma},\lambda_{0})={{\lambda_{0}(1-2r)
(1-y+1/(1-y))+\lambda_{e} r\zeta[1+(1-y)(1-2r)^{2}]}
\over{1-y+1/(1-y)-4r(1-r)-\lambda_{e}\lambda_{0}r\zeta (2r-1)(2-y)}}
\end{eqnarray}
where $E_{\gamma}$ is the energy of backscattered photons. The
differential cross section for the subprocess is then written
through the formula

\begin{eqnarray}
d\hat \sigma(\lambda^{(1)}_0,\lambda^{(2)}_0
;\lambda_{W^+},\lambda_{W^-},\lambda_Z)
=\frac{1}{4}(1-\xi_1(E^{(1)}_{\gamma},\lambda^{(1)}_{0}))
(1-\xi_2(E^{(2)}_{\gamma},\lambda^{(2)}_{0}))d\hat
\sigma(-,-;\lambda_{W^+},\lambda_{W^-},\lambda_Z)\nonumber
\\+\frac{1}{4}(1-\xi_1(E^{(1)}_{\gamma},\lambda^{(1)}_{0}))
(1+\xi_2(E^{(2)}_{\gamma},\lambda^{(2)}_{0}))d\hat
\sigma(-,+;\lambda_{W^+},\lambda_{W^-},\lambda_Z)\nonumber
\\+\frac{1}{4}(1+\xi_1(E^{(1)}_{\gamma},\lambda^{(1)}_{0}))
(1-\xi_2(E^{(2)}_{\gamma},\lambda^{(2)}_{0}))d\hat
\sigma(+,-;\lambda_{W^+},\lambda_{W^-},\lambda_Z)\nonumber\\
+\frac{1}{4}(1+\xi_1(E^{(1)}_{\gamma},\lambda^{(1)}_{0}))
(1+\xi_2(E^{(2)}_{\gamma},\lambda^{(2)}_{0}))d\hat
\sigma(+,+;\lambda_{W^+},\lambda_{W^-},\lambda_Z)\nonumber
\\
\end{eqnarray}
Here
$d\hat{\sigma}(\lambda^{(1)}_{\gamma},\lambda^{(2)}_{\gamma};\lambda_{W^+},\lambda_{W^-},\lambda_Z)$
is the helicity dependent differential cross section in the helicity
eigenstates;  $\lambda^{(i)}_{\gamma}=+,-$ and $\lambda_{V}=+,-,0$
($V=W^+,\,W^-$ or $Z$). Superscript (1) and (2) represent the
incoming gamma beams and $\xi_1(E^{(1)}_{\gamma},\lambda^{(1)}_{0})$
and $\xi_2(E^{(2)}_{\gamma},\lambda^{(2)}_{0})$ represent the
corresponding helicity distributions.

The integrated cross section for $W^{+}W^{-}Z$ production via
$\gamma\gamma$ fusion can be obtained by the following integration:

\begin{eqnarray}
d\sigma(e^+ e^- \to \gamma \gamma \to
W^{+}W^{-}Z)=\int_{z_{min}}^{z_{max}}dz \,2z \,
\int_{z^2/y_{max}}^{y_{max}}
\frac{dy}{y}f_{\gamma/e}(y)f_{\gamma/e}(z^2/y)\,d\hat{\sigma}(\gamma\gamma
\to W^{+}W^{-}Z)\nonumber
\\
\end{eqnarray}
where, $d\hat{\sigma}(\gamma\gamma \to W^{+}W^{-}Z)$ is the cross
section of the subprocess defined by (12) and center of mass energy
of the $e^+e^-$ system $\sqrt s$, is related to the center of mass
energy of the $\gamma\gamma$ system $\sqrt{\hat s}$ by $\hat s
=z^2s$. We have calculated the cross sections imposing that the
polar angles of the produced gauge bosons with the beam pipe are
larger than $10^0$.

\begin{figure}
\includegraphics{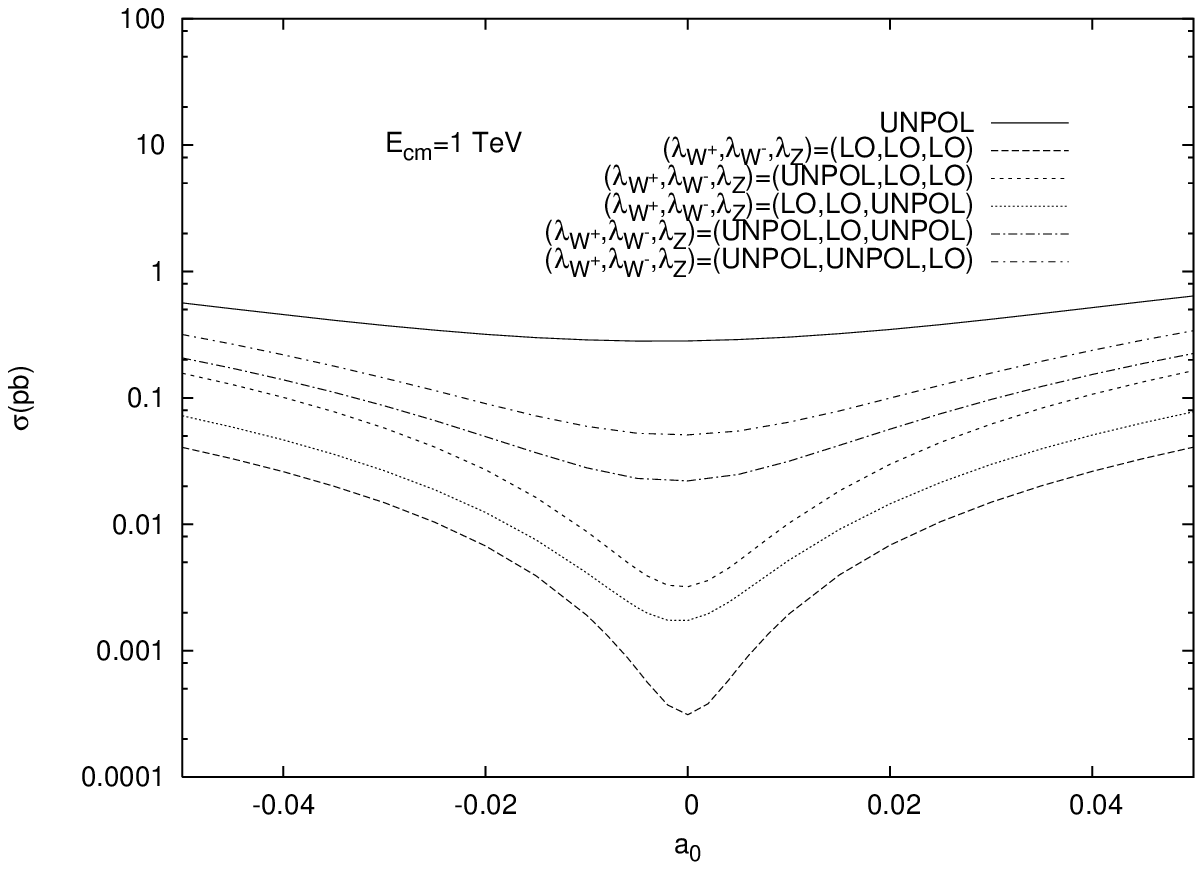}
\caption{The integrated total cross section of $\gamma\gamma \to
W^{+}W^{-}Z$ as a function of anomalous coupling $a_{0}$ for various
final state polarizations stated on the figure. Initial beams are
unpolarized and $\sqrt{s}=1$ TeV. \label{fig1}}
\end{figure}

\begin{figure}
\includegraphics{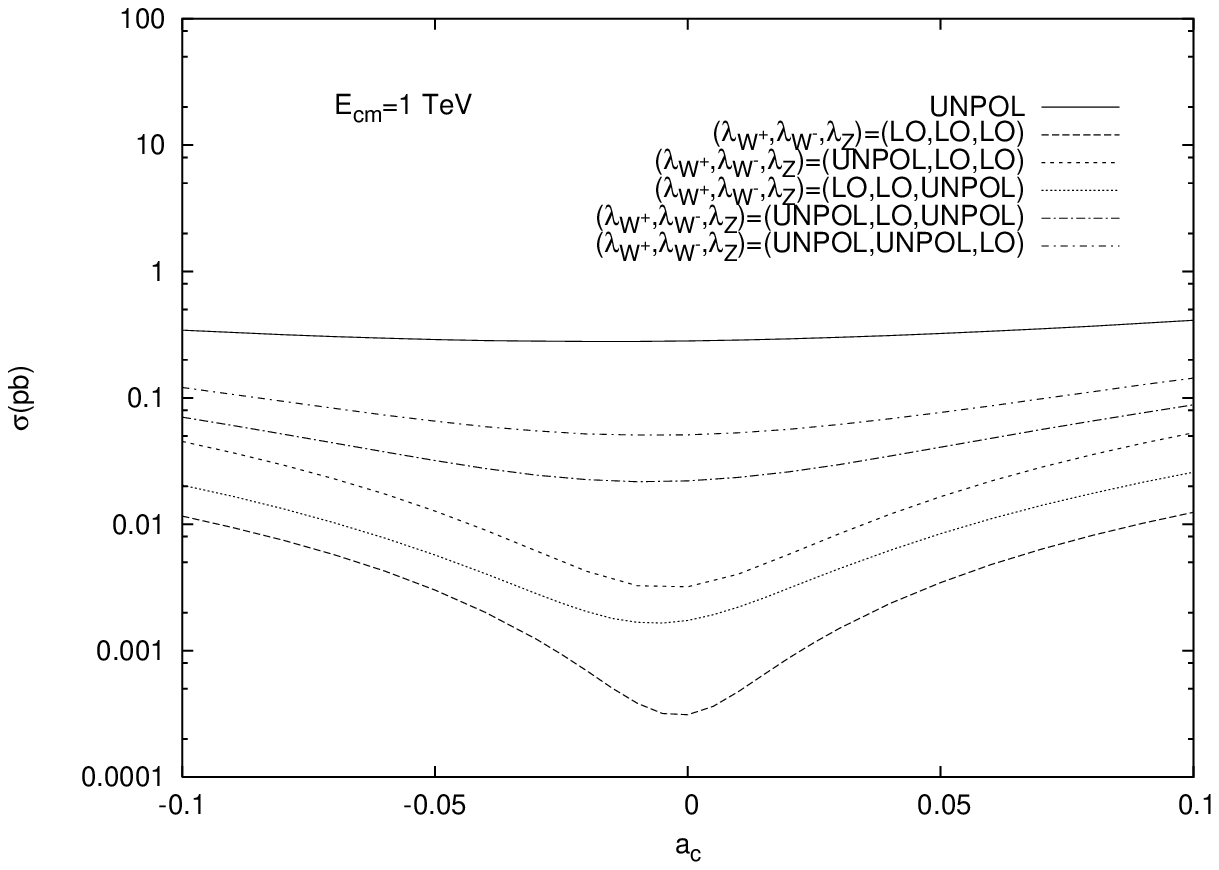}
\caption{The same as Fig.\ref{fig1} but for anomalous coupling
$a_{c}$. \label{fig2}}
\end{figure}

\begin{figure}
\includegraphics{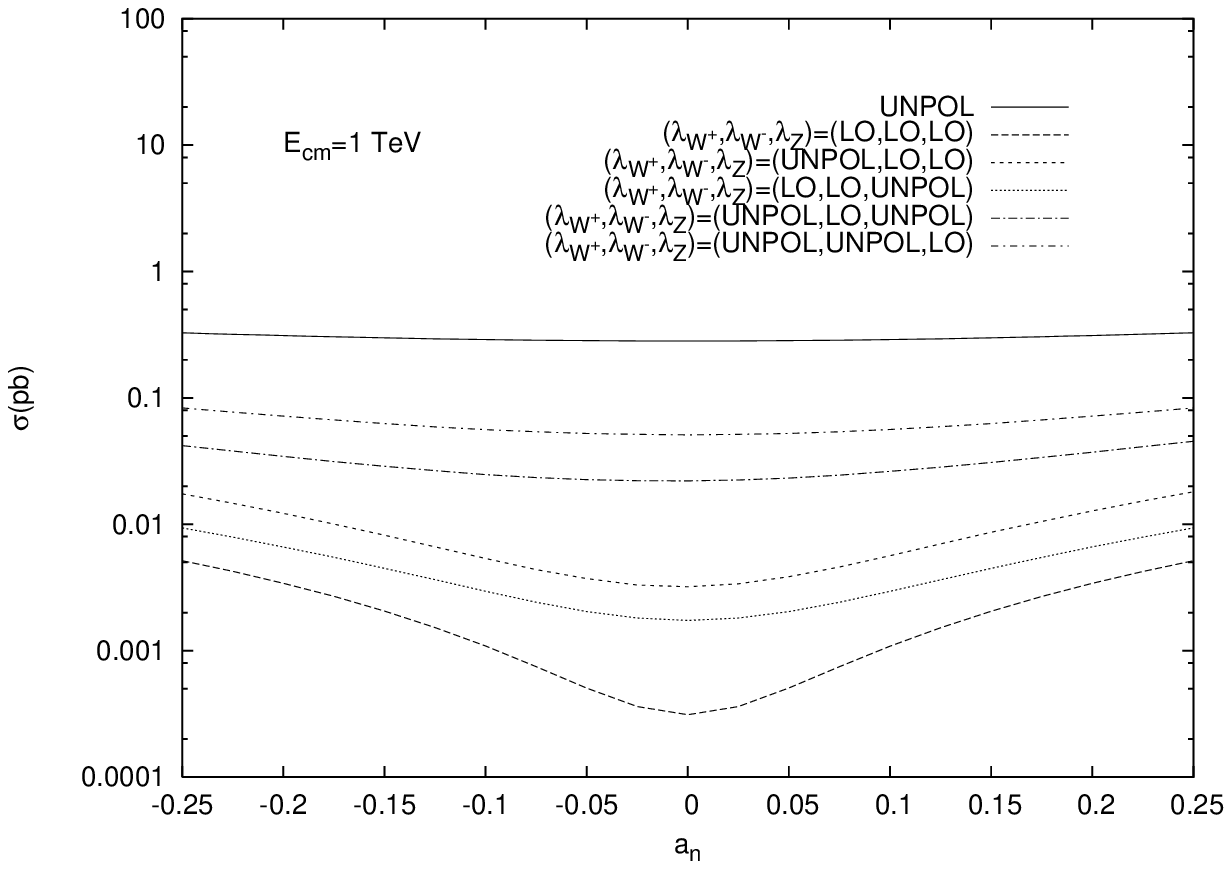}
\caption{The same as Fig.\ref{fig2} but for anomalous coupling
$a_{n}$. \label{fig3}}
\end{figure}

One can see from Fig.\ref{fig1} - \ref{fig3} the influence of the
final state polarizations on the deviations of the integrated total
cross sections from their SM value for unpolarized initial beams. In
these figures LO stands for 'longitudinal' and UNPOL stands for
'unpolarized'. Transverse polarization configurations of the final
bosons are almost insensitive to anomalous couplings. Therefore we
omit them from the figures. It is clear from figures that
longitudinally polarized cross sections are sensitive to the
anomalous couplings. The most sensitive polarization configurations
are $(\lambda_{W^+},\lambda_{W^-},\lambda_{Z})=$(LO, LO, LO) and
(UNPOL, LO, LO). For instance in Fig.\ref{fig1} the cross sections
at the polarization configurations (LO, LO, LO) and (UNPOL, LO, LO)
increase by factors of 22 and 10 as $a_0$ increases from 0 to 0.02.
But this increment is only a factor of 1.3 in the unpolarized case.

\begin{figure}
\includegraphics{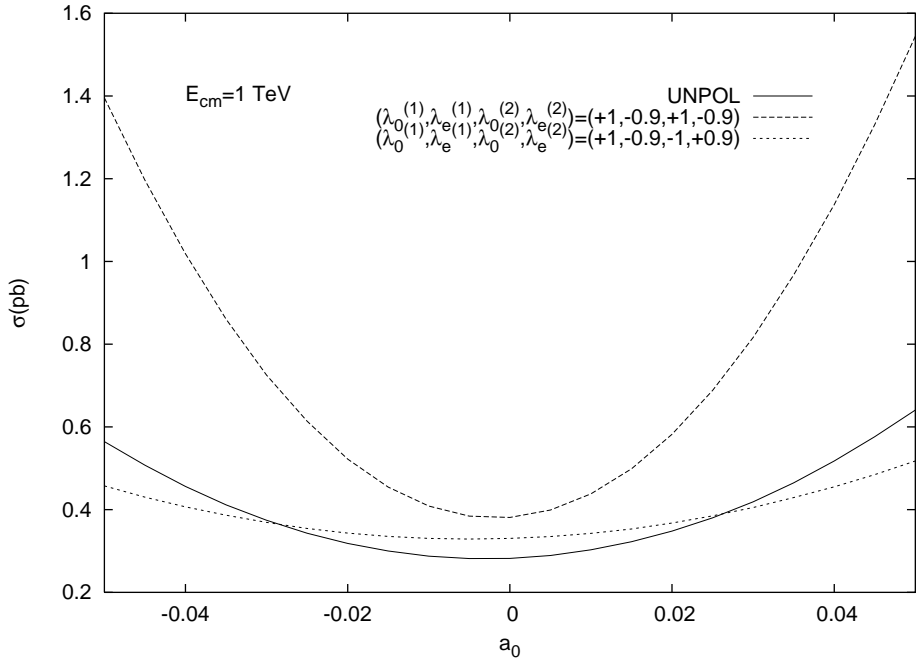}
\caption{The integrated total cross section of $\gamma\gamma \to
W^{+}W^{-}Z$ as a function of anomalous coupling $a_{0}$ for various
initial state polarizations stated on the figure. Final state gauge
bosons are unpolarized and $\sqrt{s}=1$ TeV. \label{fig4}}
\end{figure}

\begin{figure}
\includegraphics{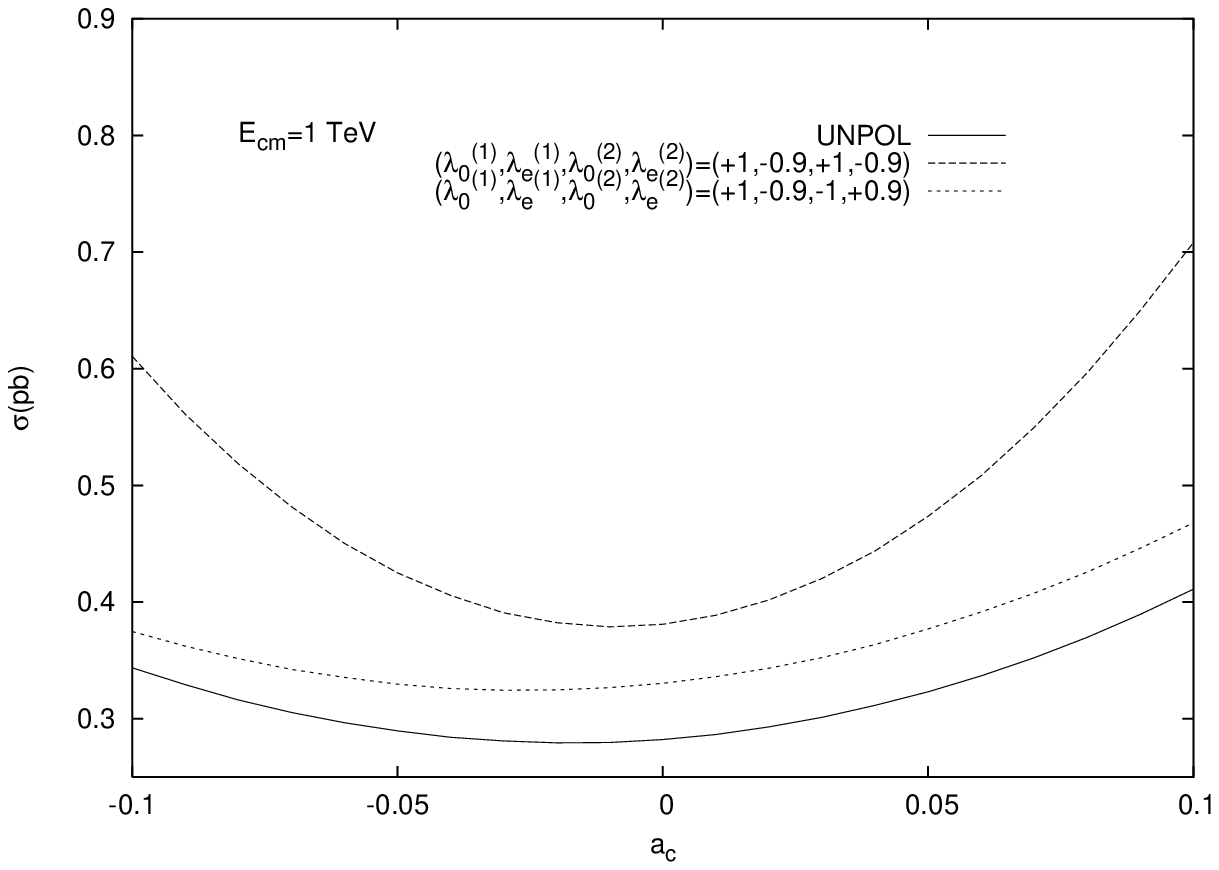}
\caption{The same as Fig.\ref{fig4} but for anomalous coupling
$a_{c}$. \label{fig5}}
\end{figure}

\begin{figure}
\includegraphics{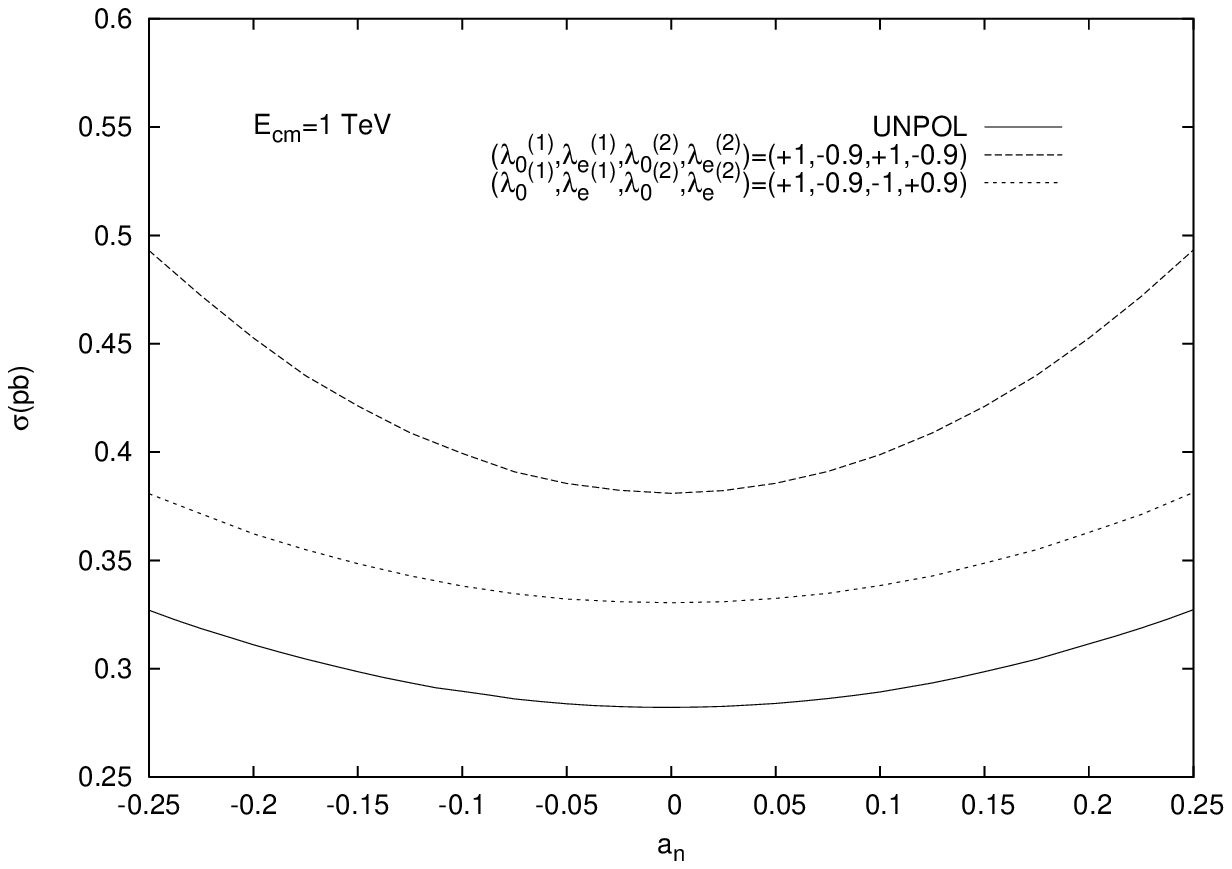}
\caption{The same as Fig.\ref{fig5} but for anomalous coupling
$a_{n}$. \label{fig6}}
\end{figure}

In Fig.\ref{fig4} - \ref{fig6} we see the influence of the initial
state polarizations on the deviations of the integrated total cross
sections from their SM value. We accept that initial electron beam
polarizability is $|\lambda_e|=0.9$. It can be shown from
backscattered photon distribution (8) that photoproduction cross
section for $\lambda_0 \lambda_e>0$ is very low at high energies
\cite{isahin2}. Therefore we have only considered the case
$\lambda_0 \lambda_e<0$ in the cross section calculations. Moreover
interchanging backscattered photon polarizations ($\xi_1
\leftrightarrow \xi_2$) do not change the cross section due to the
symmetry. During calculations we consider two different polarization
combination;
$(\lambda^{(1)}_0,\lambda^{(1)}_e,\lambda^{(2)}_0,\lambda^{(2)}_e)=$
$(+1,-0.9,+1,-0.9)$ and $(+1,-0.9,-1,+0.9)$.

\section{SENSITIVITY TO ANOMALOUS COUPLINGS}

We estimate sensitivity of the $\gamma\gamma$ collider to anomalous
couplings using simple one parameter $\chi^2$ criterion for the
integrated luminosity of $500fb^{-1}$ and $\sqrt s =0.5,\,1$ and 1.5
TeV energies. In our calculations we ignore systematic errors and
new physics energy scale $\Lambda$ is taken to be $M_{W}$.

We assume that W and Z polarizations can be measured. Indeed angular
distributions of the W and Z decay products have clear correlations
with the helicity states of them. For fixed W and Z helicities the
polarized cross sections can be obtained from a fit to polar angle
distributions of the W and Z decay products in the W and Z rest
frames. Therefore we consider the case in which W and Z momenta are
reconstructible. We restrict ourselves to leptonic decay channel of
Z boson with a branching ratio $B(Z \to \ell \bar \ell)\approx0.067$
($\ell=e$ or $\mu$) and hadronic decay channel of W boson with a
branching ratio $B(W\to q\bar{q}^\prime)\approx0.676$. The number of
events are given by $N=E B(Z \to \ell \bar \ell) B^2(W\to
q\bar{q}^\prime) L_{int}\sigma$ where $E$ is the efficiency and it
is taken to be 0.85. Leptonic decay channel of W may not be
appropriate since the momentum of the neutrino coming from the W
decay can not be determined, and consequently, the W rest frame is
unknown. For hadronic W decays, the quark charge is difficult to
reconstruct experimentally and only the absolute value of the cosine
of the decay angle can be measured. However, polar angle
distribution can still be used to measure the transverse and
longitudinal polarization states \cite{LEP4}. There have been
several experimental studies in the literature for the measurement
of W polarization. W boson polarization has been studied at CERN
$e^+ e^-$ collider LEP2 via the process $e^+ e^- \to W^+ W^- \to
\ell \nu q \bar {q}^\prime$ \cite{LEP,LEP4}. At Fermilab Tevatron,
polarization of the W bosons produced in the top quark decay has
been measured by the CDF and D0 collaborations \cite{Tevatron3}.

In Table \ref{tab1}-\ref{tab3} we show $95\%$ C.L. sensitivity
limits on the anomalous coupling parameters $a_0,\,a_c$ and $a_n$
for $\sqrt s=0.5$, 1 and 1.5 TeV energies. In the tables, LO
represents the longitudinal and UNPOL represents the unpolarized W
and Z bosons. Transverse polarization states of the final gauge
bosons are omitted since the limits for transverse polarization are
weak. We see from these tables that limits on the anomalous
couplings are usually very sensitive in
$(\lambda^{(1)}_0,\lambda^{(1)}_e,\lambda^{(2)}_0,\lambda^{(2)}_e)=(+1,-0.9,+1,-0.9)$
initial and $(\lambda_{W^+},\lambda_{W^-},\lambda_{Z})=$ (LO, LO,
LO), (UNPOL, LO, LO) final polarization configurations. For
instance, $(+1,-0.9,+1,-0.9)$ initial state polarization together
with (LO, LO, LO) final state polarization (Table \ref{tab2})
improve the lower bounds of the anomalous couplings $a_0$ and $a_c$
approximately a factor of 5 at $\sqrt s=0.5$ TeV when compared with
the unpolarized case (Table \ref{tab1}). Same polarization
configuration improve the upper bounds of $a_0$ and $a_c$
approximately factors of 4 and 3.3 at 1.5 TeV. Polarization improves
also the sensitivity limits on $a_n$. $(+1,-0.9,+1,-0.9)$ initial
state polarization together with (UNPOL, UNPOL, LO) final state
polarization improves both the upper and lower bounds of $a_n$
approximately by a factor of 2 at 0.5 TeV. The most sensitive bounds
on $a_n$ are obtained in (LO, LO, LO) at 1 and 1.5 TeV (Table
\ref{tab2}). This polarization configuration improves the limits on
$a_n$ approximately by a factor of 3 at 1 TeV and 4 at 1.5 TeV.

\section{Conclusions}

We have obtained a considerable improvement in the sensitivity
bounds by taking into account incoming beam polarizations and also
the final state polarizations of the gauge bosons. The subprocess
$\gamma \gamma \to W^{+}W^{-}Z$ in the $\gamma\gamma$ mode of a
linear collider with luminosity 500 $fb^{-1}$ probes the anomalous
$WW\gamma \gamma$, $ZZ\gamma\gamma$ and $WWZ\gamma$ couplings with a
far better sensitivity than the present collider LEP2 experiments.
It improves the sensitivity limits by up to a factor of $10^5$ with
respect to LEP2.

 It is stated in Ref.\cite{eboli2} that bounds on
the anomalous couplings $a_0$, $a_c$ and $a_n$ coming from $\gamma
\gamma \to W^{+}W^{-}Z$ are better than the ones that can be
obtained in an $e^+e^-$ collider. Our bounds on $a_0$ are a factor
from 3.5 to 10 better than the bounds obtained in $e\gamma \to Z
\gamma e$ and a factor from 1 to 5 better than the bounds obtained
in $e\gamma \to W \gamma \nu$ depending on energy \cite{isahin1}.
Our bounds on $a_c$ are approximately same with the bounds obtained
in $e\gamma \to W \gamma \nu$ and little better than the bounds
obtained in $e\gamma \to Z \gamma e$. On the other hand our bounds
on $a_n$ are little worse than the ones coming from $e\gamma \to
WZ\nu$ \cite{isahin2}.

\begin{acknowledgments}
The author acknowledges support through the Scientific and Technical
Research Council (TUBITAK) BIDEP-2219 grant.
\end{acknowledgments}

%\pagebreak

\pagebreak

\begin{table}
\caption{Sensitivity of $\gamma\gamma \to W^{+}W^{-}Z$ to anomalous
quartic $WW\gamma \gamma$ and $WWZ\gamma$ couplings at 95\% C.L. for
$\sqrt{s}$=0.5, 1, 1.5 TeV and $L_{int}=500 fb^{-1}$. Initial beams
are unpolarized. The effects of final state $W^{+},W^{-}$ and $Z$
boson polarizations are shown in each row.\label{tab1}}
\begin{ruledtabular}
\begin{tabular}{ccccccc}
$\sqrt{s}$\,\,(TeV) & $\lambda_{W^+}$ & $\lambda_{W^-}$ & $\lambda_{Z}$& $a_0$ & $a_c$ & $a_n$\\
\hline \hline
0.5& LO   & LO  & LO &-0.130, 0.139 &-0.475, 0.488 &-1.261, 1.261 \\
0.5& UNPOL & LO  & LO &-0.113, 0.075 &-0.445, 0.250 &-1.089, 0.758 \\
0.5&  LO  & LO  & UNPOL  &-0.210, 0.139 &-0.789, 0.459 &-1.441, 1.441 \\
0.5&  UNPOL  & LO  & UNPOL  &-0.174, 0.089 &-0.713, 0.269 &-1.294, 0.859 \\
0.5&  UNPOL  & UNPOL  & LO  &-0.153, 0.077 &-0.628, 0.237 &-0.878, 0.874 \\
0.5&  UNPOL  & UNPOL  & UNPOL &-0.291, 0.066 &-1.250, 0.181 &-1.076, 1.076 \\
\hline
 1& LO   & LO  & LO  &-0.0044, 0.0043 &-0.0181, 0.0143 &-0.0626, 0.0626 \\
 1& UNPOL & LO  & LO  &-0.0045, 0.0034 &-0.0194, 0.0110 &-0.0675, 0.0616 \\
 1&  LO  & LO  & UNPOL  &-0.0059, 0.0041 &-0.0258, 0.0130 & -0.0766, 0.0766\\
 1&  UNPOL  & LO  & UNPOL  &-0.0070, 0.0047 &-0.0303, 0.0148 &-0.0965, 0.0763 \\
 1&  UNPOL  & UNPOL  & LO  &-0.0071, 0.0049 &-0.0299, 0.0159 &-0.0857, 0.0860 \\
 1&  UNPOL  & UNPOL  & UNPOL  &-0.0118, 0.0060 &-0.0529, 0.0180 &-0.1127, 0.1127 \\
\hline
 1.5& LO   & LO  & LO  &-0.0008, 0.0008 &-0.0035, 0.0027 &-0.0114, 0.0114 \\
 1.5& UNPOL & LO  & LO  &-0.0009, 0.0008 &-0.0040, 0.0024 &-0.0143, 0.0145 \\
 1.5&  LO  & LO  & UNPOL  &-0.0010, 0.0008 &-0.0046, 0.0026 &-0.0149, 0.0149 \\
 1.5&  UNPOL  & LO  & UNPOL  &-0.0014, 0.0011 &-0.0059, 0.0036 &-0.0216, 0.0199 \\
 1.5&  UNPOL  & UNPOL  & LO  &-0.0015, 0.0012 &-0.0061, 0.0041 &-0.0223, 0.0226 \\
 1.5&  UNPOL  & UNPOL  & UNPOL  &-0.0023, 0.0016 &-0.0096, 0.0053 &-0.0309, 0.0309 \\
\end{tabular}
\end{ruledtabular}
\end{table}

\begin{table}
\caption{Sensitivity of $\gamma\gamma \to W^{+}W^{-}Z$ to anomalous
quartic $WW\gamma \gamma$ and $WWZ\gamma$ couplings at 95\% C.L. for
$\sqrt{s}$=0.5, 1, 1.5 TeV and $L_{int}=500 fb^{-1}$. Initial beam
polarization is
$(\lambda^{(1)}_0,\lambda^{(1)}_e,\lambda^{(2)}_0,\lambda^{(2)}_e)=(+1,-0.9,+1,-0.9)$.
The effects of final state $W^{+},W^{-}$ and $Z$ boson polarizations
are shown in each row.\label{tab2}}
\begin{ruledtabular}
\begin{tabular}{ccccccc}
$\sqrt{s}$\,\,(TeV) & $\lambda_{W^+}$ & $\lambda_{W^-}$ & $\lambda_{Z}$& $a_0$ & $a_c$ & $a_n$\\
\hline \hline
0.5& LO   & LO  & LO &-0.059, 0.064 &-0.234, 0.252 &-0.611, 0.611 \\
0.5& UNPOL & LO  & LO &-0.068, 0.036 &-0.273, 0.140 &-0.666, 0.405 \\
0.5&  LO  & LO  & UNPOL  &-0.127, 0.066 &-0.506, 0.256 &-0.950, 0.950 \\
0.5&  UNPOL  & LO  & UNPOL  &-0.112, 0.044 &-0.459, 0.168 &-0.914, 0.506 \\
0.5&  UNPOL  & UNPOL  & LO  &-0.100, 0.039 &-0.408, 0.148 &-0.549, 0.546 \\
0.5&  UNPOL  & UNPOL  & UNPOL &-0.217, 0.032 &-0.898, 0.119 &-0.737, 0.737 \\
\hline
 1& LO   & LO  & LO  &-0.0023, 0.0022 &-0.0095, 0.0082 &-0.0377, 0.0377 \\
 1& UNPOL & LO  & LO  &-0.0026, 0.0018 &-0.0109, 0.0068 & -0.0416, 0.0364\\
 1&  LO  & LO  & UNPOL  &-0.0033, 0.0021 &-0.0139, 0.0076 &-0.0502, 0.0502 \\
 1&  UNPOL  & LO  & UNPOL  &-0.0040, 0.0026 &-0.0170, 0.0097 &-0.0641, 0.0492 \\
 1&  UNPOL  & UNPOL  & LO  &-0.0041, 0.0028 &-0.0174, 0.0104 &-0.0549, 0.0549 \\
 1&  UNPOL  & UNPOL  & UNPOL &-0.0069, 0.0035 &-0.0301, 0.0126 &-0.0765, 0.0765 \\
\hline
 1.5& LO   & LO  & LO  &-0.0005, 0.0004 &-0.0019, 0.0016 &-0.0073, 0.0073 \\
 1.5& UNPOL & LO  & LO  &-0.0005, 0.0004 &-0.0022, 0.0016 &-0.0089, 0.0088 \\
 1.5&  LO  & LO  & UNPOL  &-0.0006, 0.0004 &-0.0024, 0.0016 &-0.0099, 0.0099 \\
 1.5&  UNPOL  & LO  & UNPOL  &-0.0008, 0.0006 &-0.0033, 0.0023 &-0.0141, 0.0128 \\
 1.5&  UNPOL  & UNPOL  & LO  &-0.0009, 0.0007 &-0.0036, 0.0026 &-0.0140, 0.0143 \\
 1.5&  UNPOL  & UNPOL  & UNPOL  &-0.0013, 0.0010 &-0.0055, 0.0035 &-0.0202, 0.0202 \\
\end{tabular}
\end{ruledtabular}
\end{table}

\begin{table}
\caption{The same as Table \ref{tab2} but for
$(\lambda^{(1)}_0,\lambda^{(1)}_e,\lambda^{(2)}_0,\lambda^{(2)}_e)=(+1,-0.9,-1,+0.9)$.
\label{tab3}}
\begin{ruledtabular}
\begin{tabular}{ccccccc}
$\sqrt{s}$\,\,(TeV) & $\lambda_{W^+}$ & $\lambda_{W^-}$ & $\lambda_{Z}$& $a_0$ & $a_c$ & $a_n$\\
\hline \hline
0.5& LO   & LO  & LO &-0.285, 0.302 & -0.597, 0.545 &-2.146, 2.146 \\
0.5& UNPOL & LO  & LO &-0.212, 0.162 &-0.571, 0.256 &-1.684, 1.173 \\
0.5&  LO  & LO  & UNPOL  &-0.394, 0.308 &-0.897, 0.440 &-1.335, 1.335\\
0.5&  UNPOL  & LO  & UNPOL  &-0.307, 0.190 &-0.878, 0.232 &-1.303, 0.986 \\
0.5&  UNPOL  & UNPOL  & LO  &-0.272, 0.166 &-0.782, 0.214 &-1.165, 1.160 \\
0.5&  UNPOL  & UNPOL  & UNPOL &-0.442, 0.146 &-1.462, 0.144 &-1.127, 1.127 \\
\hline
 1& LO   & LO  & LO  &-0.0067, 0.0065 &-0.0239, 0.0135 &-0.0564, 0.0564 \\
 1& UNPOL & LO  & LO  &-0.0068, 0.0051 &-0.0255, 0.0101 &-0.0690, 0.0706 \\
 1&  LO  & LO  & UNPOL  &-0.0089, 0.0061 &-0.0346, 0.0113 &-0.0619, 0.0619 \\
 1&  UNPOL  & LO  & UNPOL  &-0.0105, 0.0069 &-0.0388, 0.0134 &-0.0885, 0.0755 \\
 1&  UNPOL  & UNPOL  & LO  &-0.0107, 0.0073 &-0.0377, 0.0150 &-0.0922, 0.0924 \\
 1&  UNPOL  & UNPOL  & UNPOL  &-0.0181, 0.0087 &-0.0671, 0.0161 &-0.1103, 0.1103 \\
\hline
 1.5& LO   & LO  & LO  &-0.0012, 0.0011 &-0.0044, 0.0025 &-0.0092, 0.0092 \\
 1.5& UNPOL & LO  & LO  &-0.0013, 0.0011 &-0.0049, 0.0023 &-0.0134, 0.0146 \\
 1.5&  LO  & LO  & UNPOL  &-0.0015, 0.0011 &-0.0060, 0.0022 &-0.0115, 0.0115 \\
 1.5&  UNPOL  & LO  & UNPOL  &-0.0020, 0.0015 &-0.0072, 0.0033 &-0.0196, 0.0185 \\
 1.5&  UNPOL  & UNPOL  & LO  &-0.0021, 0.0017 &-0.0074, 0.0040 &-0.0227, 0.0231 \\
 1.5&  UNPOL  & UNPOL  & UNPOL  &-0.0033, 0.0023 &-0.0117, 0.0051 &-0.0297, 0.0297 \\
\end{tabular}
\end{ruledtabular}
\end{table}

\end{document}